\documentclass[journal,comsoc]{IEEEtran}
\usepackage[utf8]{inputenc}
\usepackage[T1]{fontenc}

\usepackage[font=scriptsize]{caption}
\usepackage{multirow}
\usepackage[pdftex]{graphicx}
\usepackage{epstopdf}
\usepackage{amsmath}
\usepackage[nocomma]{optidef}
\usepackage[colorinlistoftodos]{todonotes}
\usepackage[cmintegrals]{newtxmath}
\usepackage{xcolor}
\usepackage{cite}
\usepackage{subcaption}
\usepackage[english]{babel}

\usepackage{xspace}
\usepackage{algorithm, algorithmic} 
\usepackage{float}
\usepackage{amsfonts}
\usepackage{lipsum}
\usepackage[nocomma]{optidef}

\begin{document}
\title{\vspace{0.0cm}Ruin Theory for User Association and Energy Optimization in Multi-access Edge Computing}
\author{Do~Hyeon~Kim, Aunas~Manzoor,~
       Madyan~Alsenwi,~Yan~Kyaw~Tun,~Walid~Saad,
      ~\IEEEmembership{Fellow,~IEEE,~}
	   and~Choong~Seon~Hong,~\IEEEmembership{Senior~Member,~IEEE}
   	   \vspace{-0.5in}

\thanks{Copyright (c) 2023 IEEE. Personal use of this material is permitted. However, permission to use this material for any other purposes must be obtained from the IEEE by sending a request to pubs-permissions@ieee.org.} 
\thanks{This work was supported by the National Research Foundation of Korea(NRF) grant funded by the Korea government(MSIT) (No.2020R1A4A1018607), and part by the Institute of Information and Communications Technology Planning and Evaluation (IITP) Grant funded by the Korea Government (MSIT) (Artificial Intelligence Innovation Hub) under Grant 2021-0-02068, (No.RS-2022-00155911, Artificial Intelligence Convergence Innovation Human Resources Development (Kyung Hee University)), and (No.2019-0-01287, Evolvable Deep Learning Model Generation Platform for Edge Computing) *Dr. CS Hong is the corresponding author.} 
\thanks{Do Hyeon Kim, Aunas Manzoor, Madyan Alsenwi, Yan Kyaw Tun and Choong Seon Hong  are with the Department of Computer Science and Engineering, Kyung Hee University,  Yongin-si, Gyeonggi-do 17104, Rep. of Korea, e-mail:{\{doma, aunasmanzoor, malsenwi, ykyawtun7, cshong\}@khu.ac.kr}.}
\thanks{Walid Saad is with the Bradley Department of Electrical and Computer Engineering, Virginia Tech, Blacksburg, VA 24061 USA, and also with the Department of Computer Science and Engineering, Kyung Hee University, Yongin-si, Gyeonggi-do 17104,  Rep. of Korea, email: \ {walids@vt.edu}.}}
\markboth{ACCEPTED ARTICLE BY IEEE TRANSACTIONS ON VEHICULAR TECHNOLOGY, DOI: 10.1109/TVT.2023.3269427}%
{Shell \MakeLowercase{\textit{et al.}}: Ruin Theory for User Association and Energy Optimization in Multi-access Edge Computing}
\maketitle

\begin{abstract}
In this correspondence, a novel framework is proposed for analyzing data offloading in a multi-access edge computing system. Specifically, a two-phase algorithm, is proposed, including two key phases: \emph{1) user association phase} and \emph{2) task offloading phase}. In the first phase, a ruin theory-based approach is developed to obtain the users association considering the users' transmission reliability and resource utilization efficiency. Meanwhile, in the second phase, an optimization-based algorithm is used to optimize the data offloading process. In particular, ruin theory is used to manage the user association phase, and a ruin probability-based preference profile is considered to control the priority of proposing users. Here, ruin probability is derived by the surplus buffer space of each edge node at each time slot. Giving the association results, an optimization problem is formulated to optimize the amount of offloaded data aiming at minimizing the energy consumption of users. Simulation results show that the developed solutions guarantee system reliability, association efficiency under a tolerable value of surplus buffer size, and minimize the total energy consumption of all users.
\end{abstract}
\vspace{-0.05in}
\IEEEpeerreviewmaketitle
\vspace{-1.3em}\section{Introduction}
Multi-access edge computing (MEC) is a promising paradigm that brings various benefits to the access network in terms of communication, computing and caching \cite{MEC}. Moreover, MEC plays a key role in 5G networks in bringing cloud functionalities at the edge of the radio access network, in close proximity to end users (i.e., MEC users), thus can be derived new network paradigms, such as vehicular edge networks (VEC) \cite{9678373, 9650754}. However, resource scarcity at MEC servers is a major challenge that must be overcome in order to meet the stringent service demands of MEC users. Moreover, due to the limited resources in MEC, tasks that cannot be offloaded will be processed locally, which is also challenging as the computation capacity and battery lifetime of mobile devices are limited. Thus, users or tasks should be assigned to suitable MEC servers for improving resource utilization and ensuring system reliability of MEC server \cite{8734753, 9599379}. The authors in \cite{R3C5-3} studied the joint optimization problem of the service caching placement, computation offloading decisions, and system resource allocation. A reduced-complexity algorithm was considered for both caching placement and offloading decisions, and then an alternating minimization was further devised for balancing between system performance and computational complexity.
In \cite{R3C5-1}, a multi-user collaborative data caching and computing offloading strategy was proposed to minimize the total cost of all users in an two-tier MEC environment. More precisely, two sub-problems, resource allocation for task offloading and capacity-constrained caching backpack, were solved by a dynamic programming-based approach to reduce the total cost and offloading delay. A mobile offloading scheme, proposed in \cite{R3C5-2}, was designed based on deep reinforcement learning (DRL) to minimize the offloading latency during user mobility. Moreover, a Digital Twin (DT) edge network and an Actor-Critic based DRL framework were considered. In \cite{R3C5-4}, a dynamic DT was introduced to cover the uncertain data traffic and ununiformed resource demand and supply scenario in aerial-assisted Internet-of-Vehicles(IoV). Furthermore, two DT-driven incentives were proposed for optimal resource allocation in such system. First, CPU frequency is considered to reduce the impact of IoV network dynamics. Second, optimal CPU resource allocation is conducted to minimize task processing energy and improve service quality. Unlike the aforementioned works \cite{R3C5-1, R3C5-2, R3C5-3, R3C5-4}, we consider both users' energy minimization and the system reliability by optimizing users association and task offloading. Then, we use ruin theory with limited buffer at MEC server and convex optimization tools to solve the users association and the task offloading problems, respectively.

\vspace{-0.05em}Recently, there has been an interest in using the economic concepts of ruin theory to solve wireless problems \cite{manzoor2018ruin, htike2020ruin, manzoor2020ruin, 9322606}. In \cite{manzoor2018ruin}, ruin theory was used to model the sharing of redundant spectral resources between LTE-U and WiFi systems. The study conducted in \cite{htike2020ruin} used ruin theory to predict the minimum required resources that ensure the required QoS by each user. The authors in \cite{manzoor2020ruin} leveraged ruin theory to model energy management in unmanned aerial vehicles (UAVs). In particular, the possible number of associated users is estimated based on the surplus power of the UAVs. Furthermore, ruin theory was adopted in \cite{9322606} to obtain the probability of occurrence of rare peak age of information in augmented reality (AR) over wireless terahertz systems. To this end, the ruin theory is considered an efficient technique to model the resource insolvency or expiration of information in wireless networks. Unlike the previous works in \cite{manzoor2018ruin, htike2020ruin, manzoor2020ruin, 9322606}, we use the ruin probability to solve the user association problem in MEC-enabled wireless systems. In particular, managing buffer resources is a challenging issue to improve the task completion latency, and reliability \cite{buffer2}, \cite{buffer3}. Thus, we mainly consider estimating the ruin probability of buffer capacity in the user association phase. 
The main contribution of this paper can be summarized as follows:
\begin{itemize}
    \item To attain the efficient user association phase along with buffer management, we propose a new approach based on ruin theory. In the proposed approach, a buffer management model at the MEC server is derived from a ruin theory-based economic risk model, and, then, surplus process and ruin probability are formulated as a buffer resource utilization perspective. Finally, we design a novel preference profile based on the ruin probability of buffer resources and the total data size of the user so that the association between the MEC server and the user is formed efficiently.
    \item Moreover, the proposed approach considers the energy constraint of the associated users during the task offloading process, achieving a novel and practical framework. Specifically, we formulate an optimization problem to optimize the task offloading and computation resource allocation decisions aiming at minimizing the energy consumption of each user. The block coordinate descent (BCD) method is used to decompose the formulated problem into two sub-problems to reduce the complexity and achieve a closed-form solution.
\end{itemize}
\vspace{-1.4em}
\section{System Model}
\vspace{-0.3em}
Consider the uplink transmissions between a set $\mathcal{N}$ of $N$ MEC servers and a set $\mathcal{K}$ of $K$ mobile users. Let $\mathcal{K}_n$ be the set of users associated to a MEC server $n$. We consider each user is located within the coverage region of a number of MEC servers. In addition, each MEC server is assumed to be connected to a base station (BS) using optical fiber links and thus, the communication time between the servers and the base station is ignored. Moreover, a buffer $B_{n}$ is considered in each MEC server $n$. Let $x_{k,n}(t)$ be the association variable between a user $k$ and a MEC server $n$, where $x_{k,n}(t)=1$ if the user $k$ is associated to MEC server $n$ at time slot $t$, otherwise, 0.
\vspace{-1.5em}
\subsection{Communication Model}
\vspace{-0.03in}
In our model, the spectrum is divided into orthogonal frequencies. We consider that the available bandwidth $W_{n}$ at MEC server $n$ is equally allocated to its associated users. Thus, the achievable uplink data rate of user $k$ associated with server $n$ is given by:
\vspace{-0.12in}
\begin{equation}
    R_{k,n}(t) = x_{k,n}(t)\frac{W_{n}}{\left|\mathcal{K}_{n}\right|}\log_{2}\Big(1 + \delta_{k,n}(t)\Big),
\vspace{-0.09in}
\end{equation}
where $\left|\mathcal{K}_{n}\right|$ is the number of users associated to server $n$ and $\delta_{k,n}$ is the SNR given by:
\vspace{-0.12in}
\begin{equation}
    \delta_{k,n}(t) = \frac{P_{k}g_{k,n}(t)}{\sigma^{2}},
\label{SNR}
\vspace{-0.09in}
\end{equation}
where $P_{k}$ is the transmission power of user $k$ and
$g_{k,n}(t)$ is the channel gain between user $k$ and MEC server $n$, which can be given by $g_{k,n}= 10^{-\theta/10}$, where $\theta$ is the total path loss between user $k$ and MEC server $n$. In this work, we adopt a log-distance path loss model given by: $\theta = \theta_0 + 10 \varphi \log_{10} \frac{d}{d_0} + Y_g$, where $\theta$ is the total path loss in (dB), $\theta_0$ is the path loss at the reference distance $d_0$, $d$ is the actual distance between MEC server $n$ and user $k$, $\varphi$ is the path loss exponent, and $Y_g$ is the Rayleigh random variable for fast fading \cite{oo2016offloading}. Furthermore, we consider fixed inter-MEC interference that is captured by a constant background noise $\sigma^{2}$.
\vspace{-0.22in}
\subsection{Computation and Energy Consumption Model}
\vspace{-0.03in}

We consider that each user offloads a portion of its computation task for processing at MEC server and processes the rest locally. Let $D_{k}(t)$ be the total data size (bits) of the task of user $k$ and $\alpha_{k,n}(t)$ be 
the data bits which are offloading to MEC server. Therefore, the data processed at the user $k$ can be defined as $\beta_{k}(t) = D_{k}(t) - \alpha_{k,n}(t)$ bits. Hence, the transmission latency for offloading $\alpha_{k,n}(t)$ bits from user $k$ to the MEC server $n$ can be defined as follows:
\vspace{-0.05in}
\begin{equation}
    L_{k,n}^{\textrm{trans}}(t)=\frac{\alpha_{k,n}(t)}{R_{k,n}(t)}.\label{munir1}
\vspace{-0.05in}
\end{equation}

Based on the transmission power $P_{k}$ of user $k$ and the uplink transmission latency $L_{k,n}^{\textrm{trans}}(t)$, the energy consumption for uplink transmission from user $k$ to MEC server $n$ can be expressed as follows \cite{TE}:
\vspace{-0.05in}
\begin{equation}
    E_{k,n}^{\textrm{trans}}(t) = P_{k}L_{k,n}^{\textrm{trans}}(t).
    \label{munir2}
\vspace{-0.05in}
\end{equation}

When mobile user $k$ transmits offloading task $\alpha_{k,n}(t)$ bits to MEC server $n$, the MEC server allocates a fraction of its computational resources to user $k$. The computation latency at MEC server $n$ experienced by the user $k$ will be:
\vspace{-0.05in}
\begin{equation}
    L_{k,n}^{\textrm{comp}}(t) = \frac{\mu_{0}\alpha_{k,n}(t)}{\gamma_{k,n}},    
\vspace{-0.05in}
\end{equation}
where $\gamma_{k,n}$ is the number of CPU cycles allocated from MEC server $n$ for the computation tasks of user $k$ and $\mu_{0}$ is the required computation capacity (CPU cycles) to execute one bit of data. According to \cite{voltage} and \cite{constant}, the supply voltage is approximately linearly proportional to the allocated CPU cycles $\gamma_{k,n}$. The energy consumption for local computing at the user device is proportional to the square of supply voltage $V^{2}$\cite{voltage}. Therefore, the energy consumption at MEC server $n$ during the execution of $\alpha_{k,n}(t)$ bits offloaded from user $k$ is:
\vspace{-0.05in}
\begin{equation}
    E_{k,n}^{\textrm{comp}}(t) = \eta^{n}\gamma_{k,n}^{2}\mu_{0}\alpha_{k,n}(t),
\vspace{-0.05in}
\end{equation}
where $\eta^{n}$ is a constant that depends on the MEC server chip architecture and we set it to $10^{-28}$. Then, the local computation latency at user $k$ for executing $\beta_{k}(t)$ bits of data can be given as
\vspace{-0.05in}
\begin{equation}
    L_{k,n}^{\textrm{local}}(t) = \frac{\mu_{0}\beta_{k}(t)}{\vartheta_{k}},
\vspace{-0.05in}
\end{equation}
where $\vartheta_{k}$ is the local computational resources that user $k$ requires to execute $\beta_{k}(t)$. Therefore, the local computation time $L_{k}^{\textrm{local}}(t)$ depends mainly on $\beta_{k}(t)$ and $\vartheta_{k}$. Furthermore, the energy consumption for the local computation of user $k$ is given by
\vspace{-0.05in}
\begin{equation}
    E_{k}^{\textrm{local}}(t) = \eta^{k}\vartheta_{k}^{2}\mu_{0}\beta_{k}(t),
\vspace{-0.05in}
\end{equation}
where $\eta^{k}$ is a constant which depends on the chip architecture of the local device and the value is set to $10^{-28}$.

\vspace{-0.4cm}
\subsection{Task Queuing Model}
In the system model, we consider the task queuing at both the MEC server and user. Let $q_{k}^{\textrm{leng}}(t)$ be the queue length of user $k$, and we consider that the arrival rate at the user side $v_{k}(t)$ follows a Poisson distribution. Thus, the queue length at $t+1$ can be defined as
\vspace{-0.05in}
\begin{equation}
    q_{k}^{\textrm{leng}}(t+1) = \max\{q_{k}^{\textrm{leng}}(t) - \tau\rho_{k}(t),\ 0\}+\tau v_{k}(t),
\vspace{-0.05in}
\end{equation}
where $\max\{\cdot,\ 0\}$ indicates that the amount of transmitted data cannot be exceeded the amount of the stored data, $\tau$ is the duration of time slot, and $\rho_{k}(t)$ is the task completion rate (i.e., local processing and offloading) which can be expressed as follows:
\vspace{-0.05in}
\begin{equation}
    \rho_{k}(t) = \frac{\vartheta_{k}}{\mu_{0}} + R_{k,n}(t).
\vspace{-0.05in}
\end{equation}
where $\vartheta_{k}$ denotes the local computational resources of user $k$, $\mu_{0}$ denotes the required computation capacity and $R_{k,n}(t)$ denotes the uplink data rate of user $k$, respectively as defined previously. At the server side, we consider that each server has a queue buffer to process the offloaded tasks from the associated users. The evolution of queue length at MEC node $n$ can be expressed as follows:
\vspace{-0.05in}
\begin{equation}\label{server}
    q_{n}^{\textrm{leng}}(t+1) = \text{max}\{q_{n}^{\textrm{leng}}(t) - \frac{\tau\gamma_{k,n}}{\mu_{0}},\ 0\} + \sum_{k\in \mathcal{K}_{n}}\tau R_{k,n}(t).
\vspace{-0.08in}
\end{equation}

In (\ref{server}), the arrival rate is defined in terms of the total uplink data rate of the associated users and the departure rate is captured by the task completion rate.

In the user association phase, the channel gain depends mainly on the distance between users and MEC servers as the inter-MEC interference is fixed. Thus, users choose the closest MEC server to offload their tasks. However, the latency of task execution increases if the chosen MEC server has less available buffer size than the offloaded task. Thus, here, we consider that the channel gain and the available buffer size of the MEC server impact the user association.

For the available buffer size of MEC server, we consider the aforementioned queuing model with maximum queue length. 
Moreover, we further contemplate the possibility of lack of free buffer capacity at time $t$, and we use the bankruptcy possibility of free buffer capacity estimated by ruin theory \cite{nie_dickson_li_2011} as an important factor for the user association phase. Particularly, we build a ruin probability-based preference list that is used by the MEC server to be associated with the best user set.

We discuss our proposed framework for user association and task offloading in detail in the next section.\vspace{-1em}
\section{Proposed Framework}
\subsection{User Association Phase: A Ruin-Based Approach}
In economics, ruin theory is used for risk modeling of an insurance company \cite{ruin2}. The insurance company has an initial capital and obtains a premium from the customers. When claims are made by customers, the company must pay them. When the insurance company has a negative capital and is not available to pay the claims, a ruin occurs. 
In other words, each insurance company has its own \emph{surplus process} which is a model of accumulation of capital. Therefore, the \emph{surplus process} is defined as
\vspace{-0.08in}
\begin{equation}
\label{surplus}
    U(t) = u + ct - S(t),
\vspace{-0.08in}\end{equation}
where $u$ is the initial capital, $c$ is the premium. In addition, $S(t)=\sum_{i=1}^{N(t)}Z_{i}$ represents that the total paid claims by time $t$ under a compound Poisson process in the time interval [0, $t$]. $N(t)$ is the claim arrival process that modeled using a Poisson distribution with intensity parameter $\lambda t$, and $Z_{i}$ represents the independent identically distributed (i.i.d) amount of claim with finite mean. Based on this surplus process, each insurance company considers the probability of ruin, which is the probability of a negative surplus at a certain point. Therefore, the probability of ruin is an important metric of risk management in ruin theory, and it is defined as 
\vspace{-0.08in}
\begin{equation}
\label{Probofruin}
    \Psi(t,u) = Pr(U(t) < 0|U(0) = u).
\vspace{-0.08in}\end{equation}

In our model, a limited buffer size is considered at each MEC server, and we assume that a portion of the buffer is allocated to other tasks. To apply ruin theory for user association based on a data bit unit, we consider the existing free buffer space and task processing rate with time duration $\tau$.
In addition, the existing free buffer space and task processing rate are represented as initial capital and premium. Moreover, the ruin-theoretic concept of a ``claim'' is mapped to the total arrival data bits $\sum_{k \in \mathcal{K}_{n}}\alpha_{k,n}$. Thus, the surplus process of the buffer at time $t$ can be given by: 
\vspace{-0.05in}
\begin{equation}
\label{qsur}
    B_{n}^{\textrm{s}}(t) = B_{n}^{\textrm{s}}(0) + \frac{\gamma_{k,n}\tau}{\mu_{0}}(t) - \sum_{k \in \mathcal{K}_{n}}x_{k,n}\alpha_{k,n}(t).
\vspace{-0.05in}
\end{equation}

To get the ruin probability of a MEC server's buffer size at time $t$, we use the data processing rate with time duration $\tau$ and the total arrival data bits that associated users will offload. Thus, we define the probability of ruin at a MEC server as follows:
\vspace{-0.12in}
\begin{equation}
\label{qpro}
    \Psi\left(t, B_{n}^{\textrm{s}}\right) = Pr(B_{n}^{\textrm{s}}(t) < \epsilon_{n}|B_{n}^{\textrm{s}}(0) \neq 0),
\vspace{-0.09in}
\end{equation}
where $\epsilon_{n}$ is the tolerable size for the surplus buffer at MEC server $n$. In other words, a MEC observes the occurrence of a ruin event if the surplus buffer size is less than $\epsilon_{n}$ bytes. Therefore, a MEC server uses (\ref{qpro}) as the ruin probability in the user association phase.
The claims in the surplus process are modeled using an exponential distribution of parameter $\mu$. 
Based on the initial surplus value $B_{n}^{\textrm{s}}(0)$, the premium value $\frac{\gamma_{k,n}\tau}{\mu_{0}}$ and the claim distribution parameter $\mu$, the probability of ruin $\Psi\left(t, B_{n}^{\textrm{s}}\right)$ is determined as follows \cite{finiteruin}: 
\vspace{-0.11in}\begin{multline}
  \Psi\left(t, B_{n}^{\textrm{s}}\right)= \\\sum_{j=1}^{n} \frac{[\mu c_j\left(B_{n}^{\textrm{s}}(0)\right)]^{j-1}}{(j-1)!} e^{-\mu'c_j\left(B_{n}^{\textrm{s}}(0)\right)}  \frac{c_1(B_{n}^{\textrm{s}}(0))}{c_j\left(B_{n}^{\textrm{s}}(0)\right)},  
\vspace{-0.2in}\end{multline}
where $c_j\left(B_{n}^{\textrm{s}}(0)\right) = B_{n}^{\textrm{s}}(0) + j\frac{\gamma_{k,n}\tau}{\mu_{0}}$, and $c_1(B_{n}^{\textrm{s}}(0)) = B_{n}^{\textrm{s}}(0) +\frac{\gamma_{k,n}\tau}{\mu_{0}}$. 
\setlength{\textfloatsep}{2pt}
\begin{algorithm}[t!]
	\caption{\strut Ruin theory for user association}
	\label{alg:profit}
	\begin{algorithmic}[1]
	   \STATE{\textbf{Input}: \ $B_{n}^{\textrm{s}}$, $I_{n}^{k}$, $D_{k}$, $\Psi(t, B_{n}^{\textrm{s}})$ $\forall \; k, n$} 
	   \STATE{\textbf{Output}: \ $\boldsymbol{x}$}
	   \STATE{\textbf{Initialization} : $x_{k,n}$}
	   \STATE{\textbf{repeat:}}
	   \STATE{\textbf{for} $k \in \mathcal{K}$ \textbf{do}}
	   \STATE{\;\;propose association to $n$ based on $I_{n}^{k}$}
	   \STATE{\;\;\textbf{for} $n \in \mathcal{N}$ \textbf{do}}
	   \STATE{\;\;\; \textbf{while} $J_{k}^{n} \neq \emptyset$ or $B_{n}^{\textrm{s}} \geq \epsilon_{n}$ \textbf{do}}
	   \STATE{\;\;\;\;\;\; build $J_{k}^{n}$ from proposing users using Eq. (\ref{zeta})}
	   \STATE{\;\;\;\;\;\; accept proposing user with high priority in $J_{k}^{n}$}
	   \STATE{\;\;\;\;\;\; $B_{n}^{\textrm{s}}$ = $B_{n}^{\textrm{s}}$ - $D_{k}$}
	   \STATE{\;\;\;\;\;\; remove $k \in \mathcal{K}_{n}$ from $J_{k}^{n}$}
	   \STATE{\;\;\;\;\;\; update $x_{k,n}$}
	   \STATE{\;\;\; \textbf{end while}}
	   \STATE{\;\;remove $k \in \mathcal{K}_{n}$ from $\mathcal{K}$}
	   \STATE{\;\;update $I_{n}^{k}, \forall \; k \in \mathcal{K}$}
	   \STATE{\;\;\textbf{end for}}
	   \STATE{\textbf{end for}}
	   \STATE{\textbf{until} $\sum_{n \in \mathcal{N}}B_{n}^{\textrm{s}} \leq \epsilon_{n}$ or $\mathcal{K} = \emptyset$}
	\end{algorithmic}
	\label{Algorithm}
\end{algorithm}
\vspace{0.3em}

Finally, the factor $\zeta_{k,n}(t)$ is designed  based on the probability of ruin to capture the degree to which the total data size affects the probability of ruin at the MEC server $n$. In other words, a user that has a smaller total data size than others will have a smaller impact on the probability of ruin. Therefore, $\zeta_{n}(t)$ is given by:
\vspace{-0.3em}\begin{equation}
    \zeta_{k,n}(t) =\frac{D_{k}(t)}{\Psi\left(t, B_{n}^{\textrm{s}}\right)}.
    \label{zeta}
\vspace{-0.03in}\end{equation}

In the initial step of the Algorithm 1, the user chooses MEC server $n$ by using the achievable SNR value $\delta_{k,n}$ at line 5. In other words, the user builds a preference profile $I_{n}^{k}$ to choose an MEC server based on the high value of the achievable SNR. In this regard, the proposing users are sorted in an increasing order by using $\zeta_{n}(t)$ in the preference profile $J_{k}^{n}$ of the MEC server from line 8 to line 10. In other words, a user that has a smaller total data size than others will have a smaller impact on the probability of ruin, and thus, will get a high priority. Here, the total data size of each user is used to consider the worst-case scenario in which each user offloads all of their data to the MEC server. After MEC servers build preference profile $J_{k}^{n}$, association is executed between MEC servers and users through their preference profile until fulfilling buffer space of MEC server at line 11. The user who couldn't associate with the first priority MEC server updates the $I_{n}^{k}$ at line 16 and starts to propose with the second best MEC server. This iterative process ends when the whole buffer space is exhausted\footnote{In this time, the whole rejected users need to compute the task locally.} below $\epsilon_{n}$ or when there is no additional proposing user in the network. After guaranteeing efficient user association, data offloading must be performed in order to execute the tasks of each user.
\vspace{-1.3em}
\subsection{Task Offloading Phase: An Optimization-based Approach}
\begin{figure*}[t!]
	\centering
	\begin{minipage}[b]{0.3\textwidth}
	\centering
	\includegraphics[width=\linewidth, height=1.3in]
    {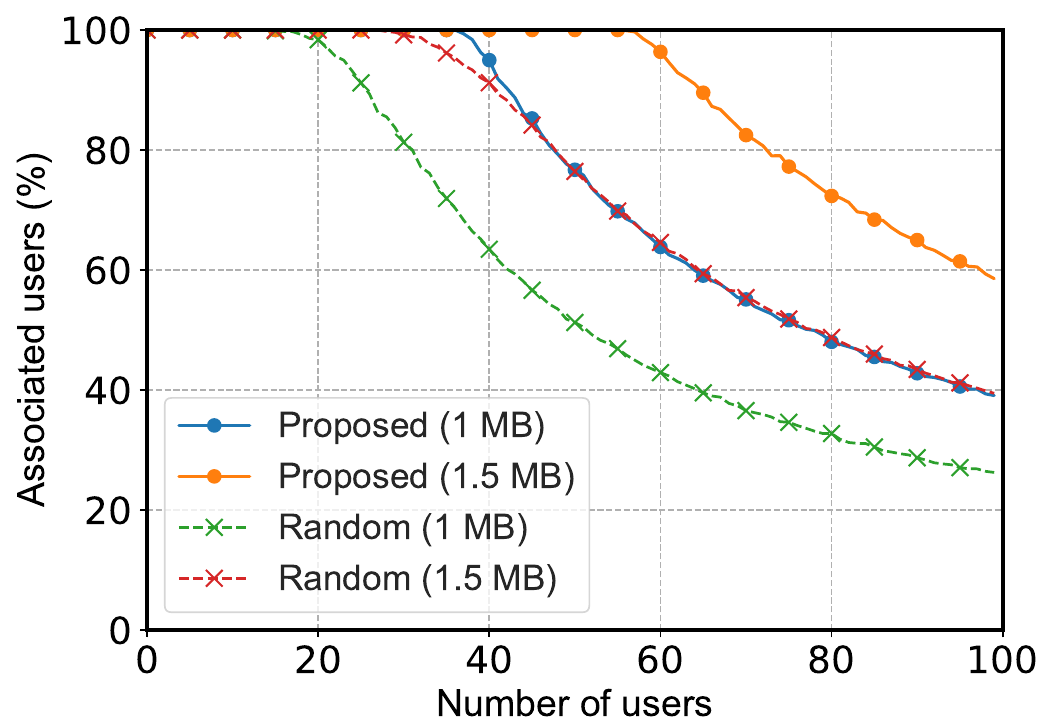}
    \captionof{figure}{Percentage of associated users for different association methods and MEC buffer sizes.}
	\label{au}
	\end{minipage}
    \hspace{0.25cm}
	\begin{minipage}[b]{0.3\textwidth}
	\includegraphics[width=\linewidth, height=1.3in]
    {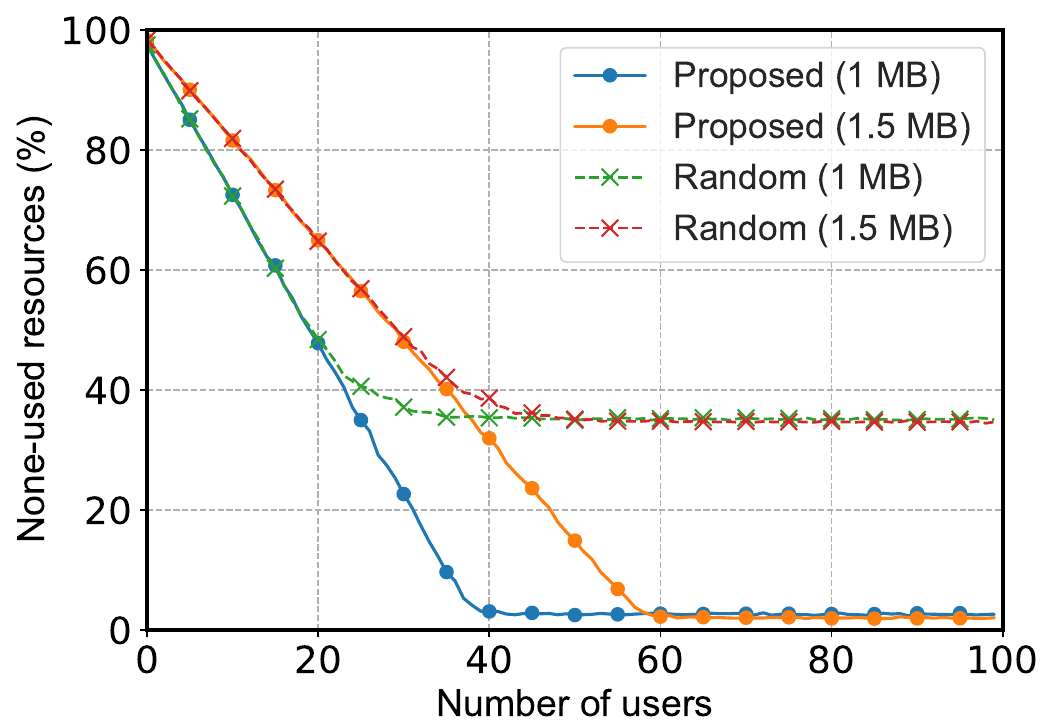}
    \captionof{figure}{Percentage of non-used resources for different association methods and MEC buffer sizes.}
	\label{nur}
	\end{minipage}
    \hspace{0.25cm}
	\begin{minipage}[b]{0.33\textwidth}
	\centering
	\includegraphics[width=\linewidth, height=1.3in]{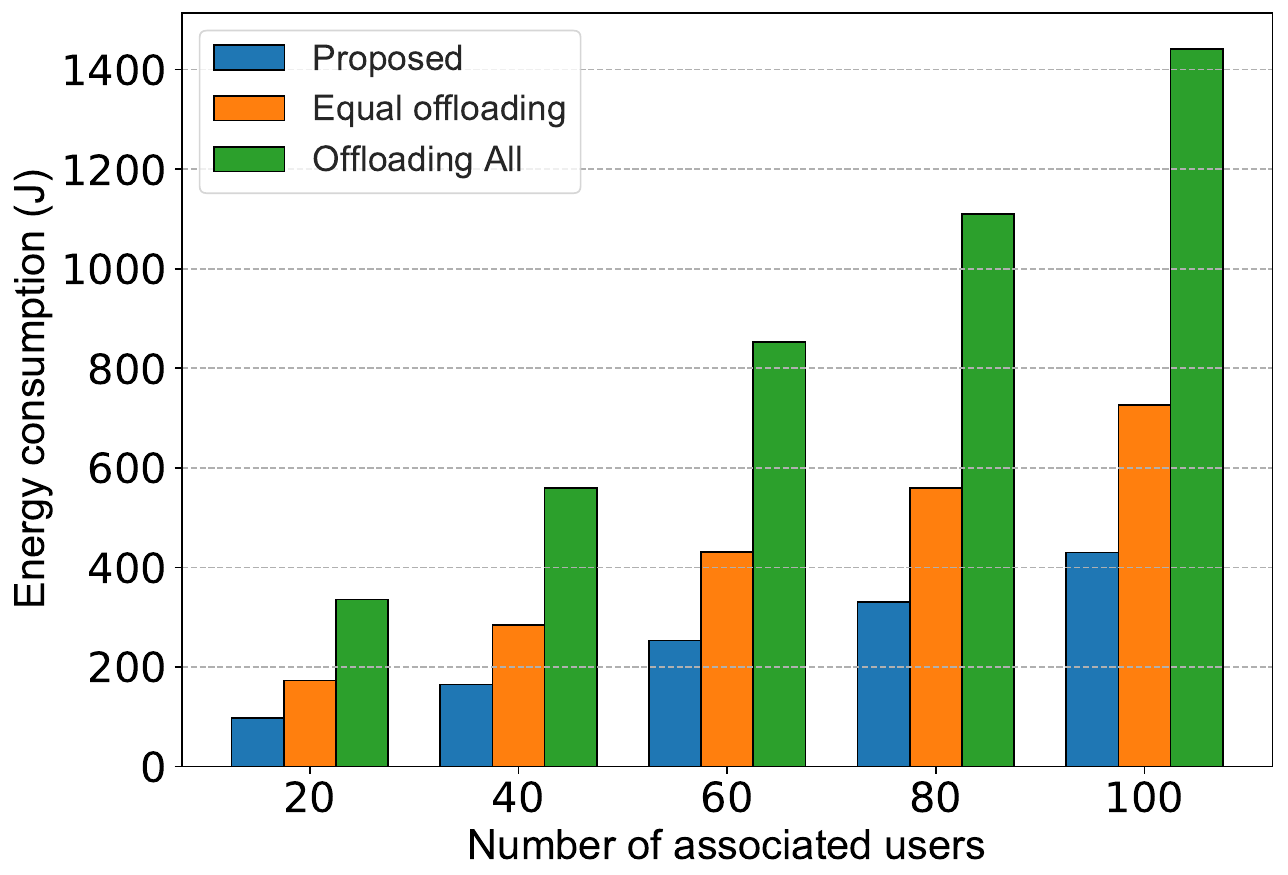}
    \captionof{figure}{Comparison for total amount of energy consumption of associated users in task offloading phase.}
	\label{optimization}
	\end{minipage}
    \hspace{0.25cm}
\end{figure*}
\vspace{-0.03in}
Our next goal is to minimize the total users' energy consumption, i.e. transmission and local computing energy consumption of the users, and the computation energy at the MEC server by optimally offloading the users' tasks. Then, we define the utility function as follows:
\vspace{-0.1in}\begin{align}
    U(\boldsymbol{\alpha})
    &= \sum_{k \in \mathcal{K}_{n}}\eta^{k}\vartheta_{k}^{2}\mu_{0}\beta_{k,n}(t)+\omega\!\sum_{k \in \mathcal{K}_{n}}P_{k}\frac{\alpha_{k,n}(t)}{R_{k,n}(t)} \notag\\
    &+ \sum_{k \in \mathcal{K}_{n}}  \eta^{n}\gamma_{k,n}^{2}\mu_{0}\alpha_{k,n}(t), \notag\\
    &= \sum_{k \in \mathcal{K}_{n}}E_{k}^{\textrm{local}}(t)+\omega\!\sum_{k \in \mathcal{K}_{n}}E_{k,n}^{\textrm{trans}}(t)+\!\sum_{k \in \mathcal{K}_{n}}E_{k,n}^{\textrm{comp}}(t)\vspace{-0.2in}
\label{utility}\end{align}
where $\omega$ is a weight to control the tradeoff between the computation energy and the transmission energy consumption. Moreover, $\beta_{k,n}$ can be replaced with $D_{k}-\alpha_{k,n}$. Based on (\ref{utility}), therefore, the optimization problem can be formulated as follows:\vspace{-0.12in}
\begin{mini!}[2]                 
	{ \boldsymbol{\alpha}, \boldsymbol{\gamma}}                               
	{U(\boldsymbol{\alpha},  \boldsymbol{\gamma} )} {\label{main_problem}}{}
 	\addConstraint{L_{k,n}^{\textrm{local}} \leq T_k, \;\forall \; k\in\mathcal{K}, \ n \in \mathcal{N},\label{const1}}
 	\addConstraint{L_{k,n}^{\textrm{comp}}+L_{k,n}^{\textrm{trans}} \leq T_k, \;\forall \; k\in\mathcal{K}, \ n \in \mathcal{N},\label{const2}}
 	\addConstraint{\sum_{k \in \mathcal{K}_{n}} \gamma_{k,n} \leq C_n, \forall n \in \mathcal{N}, \label{const4} }
 	\addConstraint{0\leq \alpha_{k,n} \leq D_k, \;\forall \; k\in\mathcal{K}, \ n \in \mathcal{N},\label{const3}}
\vspace{-0.1in}\end{mini!}

where $\boldsymbol{\alpha}$ is the task offloading vector with each element $\alpha_{k,n}(t)$ being the task of user $k$ offloaded to the server $n$ at time slot $t$, $\boldsymbol{\gamma}$ is the computation resource (i.e., CPU cycles) allocation vector with each element $\gamma_{k,n}$ being the allocated CPU cycles to compute the offloading task of user $u$, and $T_k$ is the maximum tolerable latency in order to compute the task of user $k$. (\ref{const1}) and (\ref{const2}) ensure that the tasks processing (local processing or offloading) must be completed within $T_k$. Then, (\ref{const4}) constrains the total allocated CPU cycles in order to compute the offloading task of all associated users to be smaller than the maximum available CPU cycles at the MEC server $n$. Finally, (\ref{const3}) ensures that the offloading data size is smaller than the total data size of the task. From (\ref{main_problem}), we can clearly see that the decision variables, i.e., $\boldsymbol{\gamma}$ and $\boldsymbol{\alpha}$ are coupling in both objective function (\ref{main_problem}), and the constraint (\ref{const4}). As a result, it is challenging to address the formulated optimization problem. Thus, by using the block coordinate descent (BCD) method, we decompose the formulated problem into two sub-problems. Then, the decomposed sub-problems are solved alternatively.    

At the given CPU cycles allocation at the MEC server, we can formulate the task offloading decision problem of users can be reformulated as follows:
\vspace{-0.1in}\begin{mini!}[2]                 
	{ \boldsymbol{\alpha}}                               
	{U(\boldsymbol{\alpha})} {\label{main_problem_1}}{}
    \addConstraint{(19\textrm{b}), (19\textrm{c}), (19\textrm{e})}
\vspace{-0.1in}\end{mini!}
From (\ref{main_problem_1}), we can see that the objective function and constraints are convex and linear. Therefore, we can conclude that the task offloading decision problem is a convex problem. Therefore, the problem can be solved using CVXPY toolkit. Then, at the given task offloading decision, the CPU cycles allocation problem at the MEC server can be reformulated as follows:
\vspace{-0.15in}\begin{mini!}[2]             
	{\boldsymbol{\gamma}}                   
	{U(\boldsymbol{\gamma} )} {\label{main_problem_2}}{}
    \addConstraint{(19\textrm{c}), (19\textrm{d})}
\vspace{-0.12in}\end{mini!}
Here, the objective function and constraints in (\ref{main_problem_2}) are convex and linear. Thus, the CPU cycles allocation problem in (\ref{main_problem_2}) is a convex problem. Therefore, we solve the reformulated CPU cycles allocation problem by using CVXPY toolkit.
\vspace{-1em}
\section{Simulation Results and Analysis}
We evaluate the performance of the proposed framework in extensive simulations. Statistical results are averaged over 500 independent simulation runs. 
The simulation environment consists of three BSs each having its own MEC server at each and 100 users are randomly distributed in $5000$~m $\times$ $5000$~m coverage area. In addition, each user has a computation task whose maximum total data size is set to 100 KB, and we assume that $10$ CPU cycles are needed to execute one bit of data. For each MEC server and user, we consider a computation capacity of $6\times10^5$ Hz and $7\times10^4$ Hz, respectively. The uplink transmission power of the mobile users is $200$ mW, and the system's background noise is $-174$ dBm/Hz. Finally, the available system bandwidth is considered to be $20$ MHz.    

To verify the efficiency of the proposed user association algorithm, we use a random association algorithm as a comparison baseline. In the random association method, users associate with MECs randomly without a preference profile. Fig \ref{au} shows the number of users that can be associated using different association methods and MEC buffer sizes. When the MEC buffer size is set at 1.5 MB with 80 users in the network, 73.3$\%$ of users can be associated under the proposed algorithm, while the random method guarantees only 49.5$\%$ of users can be associated. At the same time, as shown in Fig. \ref{nur}, the proposed approach ensures that 97.9$\%$ of the available resources are allocated to the users while 65.2$\%$ of resources can be utilized under the random method. Thus, the proposed user association algorithm provides efficient resource utilization and reliability under limited resource environments.

Fig. \ref{optimization} studies the performance of the proposed offloading algorithm. Here, we set $T_{k}=100$ ms. We compare the proposed offloading algorithm to \emph{1) Equal offloading:} a scheme in which all users offload half of their data to the associated server and process the remaining locally and \emph{2) Offloading all:} a scheme in which users offload all data to the associated server for processing. As shown in Fig. \ref{optimization}, the proposed approach gives the best performance in terms of energy consumption compared to other baselines. When the associated users are 100, the proposed method reduces the energy consumption by $70.05\%$ and $40.44\%$ compared with the \emph{offloading all} and \emph{equal offloading} baseline methods, respectively.
\vspace{-0.8em}
\section{Conclusion}
In this correspondence, we have proposed a ruin theory-based user association and optimized task offloading method in multi-access edge computing environment. Here, we have used the concept of a ruin probability for the MEC server's buffer size to support the user association phase. Following user association, each user can drives the amount of data that must be offloaded in order to minimize the total energy consumption of users while executing the task. The amount of offload data adjusts through the convex optimization method. Simulation results demonstrate that the proposed approach is capable of providing efficient user association in terms of buffer utilization and system reliability.  Moreover, results show that the total energy consumption of users also satisfies the minimization by using our optimization solution.
\vspace{-1em}
\nocite{*}
\bibliographystyle{IEEEtran}
\bibliography{References}
	
\end{document}